# Understanding AKT-mediated chemoresistance: the relationship between ion channels and AKT activation.

Daoudi Rédoane*

*E-mail: redoane.daoudi@unicaen.fr – University of Caen Normandy 14 000 Caen France

*Personal email address: red.daoudi@laposte.net

**Abstract**

**Overcoming chemoresistance is a challenge for multiple chemotherapeutics agents like cisplatin. ABC transporters such as MDR1 or MRPs and PI3K/AKT pathway have been proposed as actors of chemoresistance in several cancers. In this review we describe two downstream targets of Akt: c-myc and p53 in the chemoresistance. We suggest a potential link between p53, c-myc and ABC transporters expression. Consequently a link between Akt and ABC transporters-mediated chemoresistance may exist. Finally we show that Akt activation may be Orai-dependent and/or TRPC-dependent, suggesting that these ion channels could constitute a therapeutic target in cancer.**

**Keys words:** cancer; chemoresistance; chemotherapy; chemosensitivity; p53; c-myc; akt; PI3K; MDR1; MRP5; ABC transporters; ion channels; Orai; TRPC; calcium

## Introduction

Various treatment are available for patients suffering from cancer such as surgery, chemotherapy or radiation therapy. However the development of chemoresistance to therapies remains a major clinical problem. The PI3K/AKT pathway could be involved in this chemoresistance.

The class I phosphoinositide-3-kinase (PI3K) produces PtdIns (3,4,5)P3 (PIP3) from PtdIns (4,5)P2 (PIP2) by phosphorylating PIP2. PIP3 is a phospholipid that resides on the plasma membrane of cells and it activates downstream targets like the kinase Akt for example. PIP3 can be dephosphorylated by the tumor suppressor PTEN, frequently deregulated in human cancers. The pleckstrin homology domain (PH) is a protein domain of approximately 120 amino acids found in several proteins such as Akt or PDK1. This domain is able to bind phosphatidylinositol lipids within membranes like PIP3 or PIP2. When PIP3 is produced by the class I PI3K, PH domains of Akt and PDK1 bind to PIP3. PDK1 can activate Akt by phosphorylating it on threonine 308 *(1)*. Another kinase called PDK2 can activate Akt by phosphorylating it on serine 473 *(1)*. Then Akt (Akt1) regulates several downstream targets in cells and the expression of these targets depends on both cell types and cell context. By regulating its downstream targets Akt has been shown to be involved in many different cellular processes, such as regulation of glucose metabolism in insulin-response tissues, cell proliferation, cell survival, protein synthesis (via mTORC1), angiogenesis and tumorigenesis. In tumorigenesis and particularly in A549 cells (NSCLC) Akt is involved in chemoresistance to drugs like cisplatin *(2)*. In fact, several studies showed its involvement in chemoresistance *(3) (4) (5)*. In this review we focus on two downstream targets of Akt: c-myc and p53 in the chemoresistance in lung cancer but also in other cancers. We choose to mainly describe A549 cells (non small cell lung cancer) because this cell line is commonly used to model adenocarcinoma and chemoresistance in this cell line has been well documented. In a second part we explore the role of ion channels in Akt activation. Emerging evidence suggests that ion channels could activate Akt and could be involved in AKT-mediated chemoresistance in several cancers.

## The PI3K/AKT pathway and chemoresistance

### Akt, c-myc stabilization and chemoresistance

The proto-oncogene c-myc plays a role in normal cell proliferation and is involved in several cancers. C-myc is also involved in apoptosis and in chemoresistance. It has been shown that the inhibition of c-myc with c-myc siRNA improved the sensitivity of A549 cells to cisplatin *(6)*. Other studies showed that c-myc regulates ABC transporters expression such as MDR1 (or p-glycoprotein) and MRP1 *(7)*. In COLO-320 cells c-myc knockdown decreases MRP5 expression *(73)*, suggesting that c-myc activates MRP5 expression. In hepatocellular carcinoma cells high expression of MRP5 is associated with chemoresistance to cisplatin *(74)*. Furthermore, in pancreatic cancer cells, MRP5 silencing increases sensitivity to gemcitabine *(75)*. MDR1 has been shown to be responsible for the efflux of many commonly used chemotherapeutics like cisplatin in A549 cells for example *(8)*. In tumor cell line KB a study suggested a close relationship between c-myc and MDR1 expression *(9)*. Another member of the myc family, n-myc, has been shown to increase MDR1 expression in the human metastatic neuroblastoma IGR-N-91 model *(10)*. It is well known that Akt phosphorylates and inactivates GSK3b *(17)* and in his turn GSK3b is able to phosphorylate and inactivate c-myc *(11)*. GSK3b enhances threonine 58 phosphorylation and ubiquitination of c-myc *(11)* whereas Ras-activated Erks stabilize c-myc by phosphorylating it on serine 62 *(12)*. Therefore Akt could be involved in the chemoresistance in lung cancer and other cancers by stabilizing c-myc. Nevertheless the relationship between c-myc and MDR1 expression in A549 cells remains unclear.

### Akt, p53 destabilization and chemoresistance

p53 functions as a tumor suppressor and is the most frequently mutated gene in human cancers. p53 is described as "the guardian of the genome" since it is important to prevent genome mutation. This tumor suppressor can activate DNA repair, arrest cell proliferation or trigger apoptosis. Wild-type p53 is also involved in chemosensitivity in cancers. In fact, wild-type p53 decreases MRPs expression *(76)*. Moreover wild-type p53 downregulates the MDR1 promoter whereas mutant p53 upregulates MDR1 expression *(13) (19)*. In Saos-2 (human osteosarcoma) and SW620 (human colon carcinoma) cell lines another study showed that wild-type p53 represses MDR1 expression by binding to a head-to-tail (HT) site within the MDR1 promoter *(14)*. This transcriptional mechanism remains to be elucidated in A549 cells. We hypothesize that in A549 cells p53 could repress MDR1 expression because A549 cells express wild-type p53 *(15)*. Thus, in A549 cells wild-type p53 could enhance chemosensitivity to cisplatin and other drugs. However p53 is mutated in several other cell lines and certain mutations upregulate MDR1 expression *(13)*. These results strongly suggest that the p53 status need to be taken into consideration during designing therapeutic strategies targeting p53. In MCF7, Saos-2 and 293T cells it has been shown that Akt activates mdm2, an E3 ubiquitin ligase for p53, by phosphorylating it on serine 186 *(16)*. Thus, Akt enhances mdm2-mediated ubiquitination and degradation of p53 and it could induce chemoresistance in cell lines expressing wild-type p53 (like A549 cells) by destabilizing p53.

### MDR1 methylation

Surprisingly, a recent study showed that the hypermethylation of MDR1 is associated with the upregulation of MDR1 expression and chemoresistance to cisplatin in A549 cells *(8)*. This result is very interesting because it is commonly recognized that methylation of DNA represses transcription. Authors of the study hypothesize that an upstream promoter is used via the hypermethylation of the commonly used downstream promoter. This study shows that other mechanisms like epigenetic mechanisms, that seem to be p53 and c-myc independent, can regulate MDR1 expression in A549 cells. However in this review these mechanisms are not described.

Because c-myc and p53 seem to be implicated in the chemoresistance to cisplatin and other drugs we now describe the most important pathways involving c-myc, p53 and Akt in the chemoresistance.

**The PI3K/AKT, c-myc expression and beta-catenin pathways in the chemoresistance**

GSK3b is a direct downstream target of Akt *(17)*. Gsk3b has been shown to inhibit beta-catenin by phosphorylating it and promoting its degradation via the E3 ubiquitine ligase beta-TrCP *(17)*. Target genes of the beta-catenin pathway include c-myc, bcl-xl and other genes *(6) (20)*. Therefore beta-catenin-mediated chemoresistance is due, in part, to c-myc. Akt has been shown to activate this pathway either directly or indirectly. Indeed, in A431 cells (established from an epidermoid carcinoma in the skin), Akt can directly activate beta-catenin and subsequent Bcl-xl, contributing to the chemoresistance via an anti-apoptotic effect *(17) (20)*. In A549 cells, authors demonstrated that Akt indirectly activates the beta-catenin pathway by inhibiting the GSK3b kinase, resulting in Bcl-xl expression *(17)*. Thus, Akt may also directly activate beta-catenin in A549 cells.

In A549 cells overexpression of disheveled2 (DVL2) increased the protein levels of BCRP, MRP4, and survivin, all are known to be related to the chemoresistance *(18)*. It has been shown that DVL2 inhibition by its inhibitor or shDVL2 resensitizes A549 cells to cisplatin and the protein levels of BCRP, MRP4 and survivin decreased *(18)*. Inactivation of GSK3b reversed the shDVL2-induced down-regulation of beta-catenin *(18)*. This result strongly suggests that DVL2, like Akt, inhibits GSK3b to upregulate beta-catenin in A549 cells. Moreover in the canonical wnt/beta-catenin pathway disheveled is known to inhibit GSK3b in order to avoid the degradation of beta-catenin. In other cell lines, like PC12, Int5, Tni3 and 293T cells, a link between Akt and DVL pathways is well established *(21)*. This remains to be shown in lung cancer and particularly in A549 cells.

In a recent study authors highlighted a direct interaction between fibronectin 1 and integrin-beta 1 in A549 cells *(22)*. More generally, this interaction is well characterized in a wide range of cell types because it is important to mediate cell-matrix interaction. Ligands for integrins include fibronectin but also other proteins such as vitronectin, collagen and laminin. Interestingly fibronectin 1 silencing suppressed the wnt/beta-catenin pathway and knockdown of fibronectin 1 sensitized cells to cisplatin *(22)*. Fibronectin 1 was also upregulated by cisplatin itself in H1299 cells *(22)*, suggesting an acquired chemoresistance to cisplatin in these cells. These results show a functional link between fibronectin 1 and beta-catenin pathway. It is reported that integrin-beta 1 stimulates beta-catenin signaling by phosphorylating and inactivating GSK3b *(23)*. The chemoresistance to cisplatin could be explained by a direct interaction between fibronectin 1 and integrin beta-1. Then integrin beta-1, in his turn, stimulates chemoresistance to cisplatin mediated by the beta-catenin pathway, by inactivating GSK3b via integrin-linked kinase *(23)*. ILK is a conserved intracellular kinase-like protein that interacts with the cytoplasmic domain of beta-1 integrin *(29)*. ILK can be activated when an extracellular signal mediated by fibronectin 1 and integrin beta 1 interaction. ILK is able to phosphorylate and inactivate GSK3b to activate beta-catenin pathway in-vivo *(23)*. Is there a link between Akt and fibronectin pathways in A549 cells? A recent work shows that the first type III domain of fibronectin (called FnIII-1c) stimulates a PI3K/Akt signaling pathway leading to avb5 integrin activation and TRAIL resistance in NCI-H460 cells *(24)*. Moreover, there is some evidence that the ILK can phosphorylate serine 473 of Akt, like PDK2 *(1) (25)*. This result suggests a potential link between Akt and integrin-mediated chemoresistance in A549 cells. The involvement of fibronectin-1 in chemoresistance in NSCLC shows that the tumor microenvironment (i.e extracellular matrix here) plays an important role in tumorigenesis.

**The PI3K/AKT, c-myc expression and Notch pathways in the chemoresistance**

Is there a link between Akt and Notch pathways in A549 cells? It has been shown that jagged1 is a beta-catenin target gene *(26)*. Jagged1 is a canonical ligand for Notch receptors expressed by mammalian cells. For this reason we hypothesize that Akt may indirectly enhance jagged1 expression and activate Notch pathway in A549 cells. Furthermore Notch pathway has been shown to be involved in the chemoresistance to cisplatin and other drugs in several cancers such as lung, ovarian, bladder and testicular cancers *(27)*. The same review shows evidence suggesting that the Notch signaling pathway is frequently deregulated in human cancers with overexpression of Notch receptors and their ligands, consistent with our previous hypothesis *(27)*. In fact, the overexpression of the ligands of Notch receptors (particularly jagged1) could be due, in part, to activation of Akt pathway. Moreover, Akt has been reported to stimulate Notch activity in some experimental models and this is consistent with our previous hypothesis *(28)*. A reciprocal relationship between Akt and Notch pathways has been suggested because Akt is a Notch target gene. Another study showed a relationship between Akt and Notch pathways in T-ALL cells in vitro *(30)*. Pharmacological inhibition of Akt pathway using a PI3K inhibitor (LY294002) showed decreased expression of Notch-1 protein, a Notch pathway target gene *(30)*. However no changes were found in the Notch1 gene expression *(30)*. This result suggests that the link between Akt and Notch pathways is very complex and it may depend on the cell line used. C-myc is a target of several cell signaling networks including RAS/MAPK, Akt, Hedgehog, wnt/beta-catenin and Notch pathways. The latter is very interesting in this part of the review because if Notch1 gene expression is not decreased by Akt inhibition in A549 cells, c-myc gene expression may also remain unchanged. As c-myc has been related to the chemoresistance in A549 cells *(6)*, Notch-mediated chemoresistance via c-myc may remain unchanged upon Akt inhibition. In this case we hypothesize that the Notch-mediated chemoresistance via c-myc would be negligible in A549 cells because several studies show that inhibition of Akt improves the sensitivity of A549 cells to cisplatin *(2) (5)*. Thus, these pathways remain to be elucidated in A549 cells.

**PI3K/AKT, c-myc expression and Hedgehog pathways in the chemoresistance**

The Sonic Hedgehog pathway (SHH) is reactivated in A549 cells *(31)*. A recent cohort study showed that the SHH pathway is associated with chemoresistance to platinum-based chemotherapy in NSCLC *(31)*. Gli2, a transcription factor in the SHH pathway, was overexpressed in patients with a reduced overall survival *(31)*. It has been shown that Akt inhibits PKA and in his turn PKA can inactivate Gli2 *(32) (33) (34)*. The phosphorylation of Gli2 by PKA enhances Gli2 sumoylation in-vitro and in-vivo and sumoylation inhibits Gli2 transcriptional activity *(32)*. Consequently PKA inhibits the SHH pathway whereas Akt stimulates this pathway. PKA can also inhibit Gli2 by phosphorylating it on four residues located on the carboxy-terminal side of the DNA binding zinc-finger domain *(33)*. The phosphorylation leads to the degradation of Gli2 by the proteasome pathway *(34)*. Another study confirms the role of Gli2 in chemoresistance to the BET bromodomain inhibitor JQ1 in pancreatic cancer cells *(35)*. In this study authors used a siRNA Gli2. Gli2 knockdown resulted in about 50% reduction in both c-myc expression and protein levels. Moreover downregulating Gli2 resensitized cells to JQ1. Nevertheless this study doesn't describe a relationship between c-myc and ABC transporters. These results show that c-myc is upregulated by Gli2 and constitutes a SHH target gene. Because AKT stimulates the SHH pathway by inhibiting PKA, it may indirectly enhance c-myc expression in A549 cells and subsequent chemoresistance to cisplatin.

**PI3K/AKT, c-myc expression and TGF-beta pathways in the chemoresistance**

The role of the TGF beta pathway in carcinogenesis is very complex because the pathway functions as either a tumor suppressor or a pro-oncogenic pathway *(36) (37)*. TGF-beta is a tumor suppressor during early stages of tumor progression whereas it acts as a proto-oncogene in advanced cancers *(37)*. It has been shown that Akt can inhibit SMAD3 *(38)*. SMADs are proteins transducing extracellular signals from TGF-beta to the nucleus. SMAD3 inhibits the expression of c-myc *(57) (64)*. Thus, this result suggests that Akt may indirectly stimulate c-myc expression in TGF-beta pathway. However the relationship between SMAD3 and c-myc expression remains unclear in A549 cells. A study shows that the TGF beta receptor 1 is directly targeted by the microRNA miR-181b *(39)*. This study also shows that miR-181b enhances chemosensitivity of A549 cells to cisplatin in vitro and in vivo and inhibits cell proliferation in vitro. At first this result may seem paradoxical because TGF beta activates its downstream target SMAD3 via the activation of the TGF beta receptor 1. Then SMAD3 inhibits c-myc expression and enhances chemosensitivity to cisplatin. For this reason we expect that miR-181b, by inhibiting the TGF beta receptor 1, should inhibit SMAD3 activation, resulting in c-myc expression and subsequent chemoresistance. Nevertheless the chemosensitivity to cisplatin with miR-181b can be explained by the fact that TGF beta signaling also activates Akt and not only SMAD3. In fact the same study shows that miR-181b inhibitors elevate the expressions of p-Akt in A549 cells. Finally Akt may inhibit SMAD3 in A549 cells with an increased c-myc expression. This activation of Akt by the TGF beta signaling may be due to the microRNAs miR-216a and miR-217, both of which target PTEN *(40)*. These two microRNAs are induced by TGF beta in glomerular mesangial cells *(40)*. A similar mechanism may exist in A549 cells. We hypothesize that in the previous study *(39)* the pathway TGF beta/Akt/SMAD3/c-myc is dominant compared to the TGF beta/SMAD3/c-myc signaling. Authors said that in their study miR-181b functions as a tumor suppressor in A549 cells, consistent with our hypothesis. Other studies showed that this microRNA functions as either an oncogene or a tumor suppressor in different cancers. miR-181b induces cell proliferation in ovarian and cervical cancers *(41) (42)*. Furthermore miR-181b functions as an oncogene in early-stage breast cancer *(43)*. This result is consistent with the fact that TGF-beta acts as a tumor suppressor during early stages of tumor progression because TGF beta signaling is inactivated by the microRNA miR-181b (by binding to the 3'-UTR of TGF beta receptor 1) *(39)*.

Another microRNA, miR-10a, has been shown to be involved in chemoresistance to cisplatin in A549 cells because miR-10a silencing increased cisplatin sensitivity in these cells *(44)*. MiR-10a silencing also decreased MDR1 and MRP1 expression in A549 cells *(44)*, suggesting that this microRNA enhances the drug efflux of A549 cells via MDR1 and MRP1. Moreover miR-10a silencing suppressed the secretion of TGF beta in A549 cells *(44)*. Thus miR-10a is involved in the secretion of TGF beta in A549 cells. Taken together, these result suggest that miR-10a could enhance chemoresistance to cisplatin by stimulating the TGF beta/Akt/SMAD3/c-myc pathway in A549 cells. The role of Akt and c-myc is not described in this study. Finally miR-181b and miR-10a seem to have opposite effects in A549 cells.

These results confirm that microRNAs play a crucial role in cancer development.

**PI3K/AKT, p53 destabilization and HSP27 pathway in the chemoresistance**

In A549 cells p53 and HSP27 proteins have been shown to be PI3K dependent *(45)*. This result confirms the link between Akt, mdm2 and p53 described in the introduction of this review. HSP27 is a chaperone that can inhibit apoptosis by activating mdm2. The activation of mdm2 by HSP27 has been shown in HEK293A cells but not in A549 cells *(46)*. HSP27 may be involved in chemoresistance by destabilizing p53 in A549 cells (see the potential interaction between p53 and MDR1, MRPs transporters in the introduction of this review). HSP27 knockdown increases cytoplasmic p21 and cisplatin sensitivity in ovarian carcinoma cells *(47)*, suggesting that HSP27 inactivates p53 because p21 is a target of p53 and p21 is activated by p53. In another study HSP27 inhibition with OGX-427 sensitized A549 cells to erlotinib, pemetrexed, gemcitabine, cisplatin and paclitaxel *(48)*. In this study erlotinib activates HSP27 expression by stimulating the transcription factor HSF1 in A549 and HCC827 cells. HSF1 silencing by using siRNA HSF1 prevented erlotinib-induced activation of HSP27. Thus, HSF1 activates HSP27 in A549 cells. It is interesting to notice that mTORC1 activates HSF1 in HeLa cells *(49)*. This activation is not demonstrated in A549 cells. It is well known that Akt can activate its downstream target mTORC1 by inhibiting TSC2 *(50)*. Finally Per2 knockdown by shRNA protects A549 cells from cisplatin-induced apoptosis *(51)*. Moreover there was an activation of the PI3K/AKT/mTORC1 pathway. This suggests that the chemoresistance upon Per2 knockdown could be due to the activation of HSP27 via mTORC1 in A549 cells. In A549 cells these result suggest that Akt may indirectly activate HSP27 and increase HSP27-mediated chemoresistance in A549 cells.

**PI3K/AKT, p53 and autophagy in the chemoresistance**

The relationship between autophagy and chemoresistance and between p53 and autophagy is very complex *(52)*. But in a recent study there is an evidence that autophagy participates in chemoresistance to cisplatin in A549 cells *(53)*. p53 can stimulate or inhibit autophagy *(52)*. Because Akt destabilizes p53 by directly activating mdm2 or indirectly (via HSP27, see above), autophagy could be increased if p53 inhibits it in A549 cells or decreased if p53 increases it in A549 cells.

In this review we show that the stabilization of c-myc, the destabilization of p53 and the expressions of c-myc may play an important role in the chemoresistance to drugs in lung cancer and particularly in A549 cells because these cells express wild-type p53 *(15)*.

# Calcium channels and chemoresistance

In this review Akt appears to play a major role in the chemoresistance to drugs in lung cancer and other cancers because it seems to be involved in several pathways related to the chemoresistance. For this reason it is interesting to know how Akt is activated in these cells. In the second part of this review we describe the link between ion channels and the activation of Akt in several cell lines.

## Orai channels

Orai channels are calcium selective ion channels. They contain four transmembrane domains and form hexamers. Three isoforms of Orai (Orai1, Orai2 and Orai3) are found in mammalian cells. These channels are activated upon the depletion of intracellular calcium stores. This depletion (i.e decreased calcium concentration in the endoplasmic reticulum) is sensed by the STIM1 protein that relocates near the plasma membrane in order to activate Orai channels. Then calcium enters the cell, this is the store operated calcium entry (SOCE). SOCE refills intracellular calcium concentration (i.e in the endoplasmic reticulum) and is used to activate a specific activity in cell, for example proliferation. However, non-SOCE pathways have also been described in several cell lines.

**Are Orai3 calcium channels the upstream activators of Akt?**

Interestingly Ay. et al. showed that Orai3 silencing decreased Akt phosphorylation levels in NCI-H23 and NCI-H460 cells *(54)*. This result suggests that Orai3 channels are required to activate Akt and consequently Akt-mediated chemoresistance. The relationship between Orai3 channels and Akt phosphorylation remains to be elucidated in A549 cells. In tumor tissues and in the MCF7 cancer cell line Orai3 down-regulation reduced both expression and activity of c-myc *(55)*. This result suggests a functional link between Orai3 channels and c-myc. Based on the previous studies we hypothesize that calcium entering the cell through Orai3 channels activates Akt. Then Akt activates several downstream targets and is involved in several pathways described in this review. For example Akt indirectly activates c-myc by enhancing its expression and/or its stability. The link between Orai3 channels and c-myc remains to be explored in A549 cells. Because Orai3 channels seem to be involved in Akt and c-myc activation and possibly in Akt-mediated chemoresistance in lung cancer and other cancers we postulate that drugs could enhance Orai3 expression, leading to acquired chemoresistance. For example, in a previous study, erlotinib induced HSP27 activation *(48)* but HSP27 is activated by HSF1 in A549 cells *(48)* and HSF1 could be activated by mTORC1 in A549 cells. Thus, Akt activates mTORC1 and the activation of Akt is possibly mediated by Orai3 channels. Orai3 expression may be increased by erlotinib itself leading to acquired chemoresistance via erlotinib-induced activation of HSP27. Moreover it has been shown that cisplatin activates Erk1/2 and Akt in A549 cells *(56)*. Cisplatin, like erlotinib, could activate Akt by enhancing Orai3 expression in A549 cells, leading to acquired chemoresistance.

The last issue is: "if drugs increase Orai3 expression, how these channels are then activated in A549 cells?" In fact, Orai3 channels are store operated channels (SOC) activated by the depletion of internal calcium stores (in the endoplasmic reticulum). The decreased calcium concentration in the endoplasmic reticulum is sensed by the STIM1 protein. Then this protein clusters and relocates near the plasma membrane to activate Orai3 channels and other Orai channels (i.e Orai1 and/or Orai2 isoforms). For this reason a signal is necessary to activate this depletion and Orai3 channels in A549 cells. We propose three hypotheses to explain the activation of Orai3 channels in A549 cells. Firstly, it has been shown that A549 cells express all five subtypes of muscarinic receptors *(58)*. Furthermore A549 cells can release acetylcholine *(59)*. M3 receptors could be stimulated, resulting in the production of IP3 and DAG second messengers. Then, IP3 binds to the IP3 receptor on the endoplasmic reticulum and activates calcium release from the ER lumen to the cytoplasm in order to activate Orai3 channels (SOCE pathway). Secondly, ATP is frequently found in the tumor microenvironment. ATP could activate P2Y receptors expressed by NCI-H23 cells *(77)*, resulting in the production of IP3 and DAG second messengers and Orai3 activation (SOCE pathway). Then, a study identified a STIM-independent mechanism of Orai1 activation in MCF7 cells *(60)*. In these cells the study demonstrated that SPCA2 (a protein located in the golgi) constitutively activates Orai1 independently of STIM1 and store depletion (non-SOCE pathway). We hypothesize that a similar mechanism exists in A549 cells with a constitutive activation of Orai3 channels. Finally, drugs (like erlotinib or cisplatin for example) may activate Orai3 channels by unknown interactions.

**PI3K/AKT, Orai1 calcium channels and autophagy in the chemoresistance**

Autophagy seems to be involved in chemoresistance to cisplatin in A549 cells *(53)* whereas it seems to be involved in chemosensitivity in HepG2 hepatocarcinoma cells *(65)*. In fact, this study shows that 5-fluorouracil (5-FU) induces cell death by increasing autophagy. Authors show that 5-FU induces autophagy by inhibiting the PI3K/AKT/mTOR pathway and it is known that mTOR inhibits autophagy *(67)*. Furthermore the study shows that 5-FU decreases SOCE and Orai1 expression but has no effects on STIM1 and TRPC1 expressions. These results strongly suggest that Orai1 channels, like Orai3, may be the upstream activators of the PI3K/AKT pathway in HepG2 hepatocarcinoma cells. For this reason

Orai1 channels may inhibit autophagy in HepG2 hepatocarcinoma cells by activating the PI3K/AKT/mTOR pathway, resulting in increased chemoresistance to 5-FU in HepG2 hepatocarcinoma cells. We previously hypothesized that drugs like cisplatin may increase Orai3 expression in A549 cells, leading to acquired chemoresistance. In contrast, this study shows that 5-FU decreases Orai1 expression. This result suggests that Orai1 channels are not involved in acquired chemoresistance to 5-FU in HepG2 cells. This result also shows that Orai-mediated acquired chemoresistance, if it exists, is not induced by all drugs. Finally, the same study shows that Orai1 expression is increased in hepatocarcinoma tissues compared with non-cancerous tissue. This result suggests that Orai1 channels are involved in innate chemoresistance to 5-FU in HepG2 cells because Orai1 inhibits autophagy in HepG2 cells.

Interestingly another study shows very similar mechanisms in prostate cancer cell lines PC3 and DU145 *(66)*. In fact, this study shows that resveratrol (RSV) decreases both STIM1 expression and STIM1 association with TRPC1 and Orai1 channels, resulting in reduced SOCE. RSV also decreases Akt1 and mTOR activation and induces autophagic cell death in PC3 and DU145 cells. This result shows that, similar to HepG2 cells, autophagy is associated with chemosensitivity in PC3 and DU145 cells. Furthermore this result suggests that RSV induces autophagic cell death by decreasing SOCE, Akt activation and subsequent mTOR activation. In PC3 and DU145 cells, STIM1/Orai1 and STIM1/TRPC1 pathways could be the upstream activators of Akt1. Moreover, Orai1 and TRPC1 channels could inhibit autophagy via mTOR activation. For this reason these channels seem to be associated with chemoresistance in PC3 and DU145 cells. Similar to HepG2 cells and because RSV decreases STIM1 expression and SOCE, Orai1 channels are not involved in acquired chemoresistance to RSV in PC3 and DU145 cells. The study shows that STIM1 expression is increased in PC3 and DU145 cells compared to the control prostate cell line RWPE1 (non-cancerous tissue). Similar to HepG2 cells, this result suggests that STIM1 is involved in innate chemoresistance to RSV by enhancing the Akt/mTOR pathway activation, resulting in decreased autophagic cell death. However no change in TRPC1 or Orai1 expression levels were observed in this study, compared to HepG2 cells in which Orai1 was overexpressed in hepatocarcinoma tissues compared with non-cancerous tissue.

Because TRPC1 channels seem to be involved in chemoresistance to RSV in prostate cancer, we now explore the role of other TRPC channels in chemoresistance.

## TRPC channels

Transient receptor potential channels (TRP channels) are non-selective calcium-permeable channels because they are permeable to cations including sodium, calcium and magnesium. TRP channels consist of six transmembrane domains with intracellular N- and C- termini. They are activated by several stimuli including ligand-gating (taste) or temperature for example. TRP channels are widely expressed in many tissues. These channels are divided into seven sub-families: TRPC (C for canonical), TRPV (V for vanilloid), TRPA (A for ankyrin), TRPM (M for melastatin), TRPP (P for polycystin), TRPML (ML for mucolipin) and TRPN (N for NOMPC, no mechanoreceptor potential C, not found in mammals).

### PI3K/AKT, TRPC5 and TRPC6 channels in the chemoresistance

A study reports that adriamycin-resistant human breast cancer cells (MCF-7/ADM) have TRPC5-containing extracellular vesicles (EVs) on their cell surface *(68)*. Furthermore this study shows that this chemoresistance can be transferred to chemosensitive cells. This first result indicates that TRPC5 is involved in chemoresistance in MCF-7/ADM cells. In another study silencing TRPC5 expression and activity reduced MDR1 expression in MCF-7/ADM cells and it was associated with an increased chemosensitivity to adriamycin *(69)*. This result confirms the involvement of TRPC5 in chemoresistance to adriamycin in MCF-7/ADM cells. In this work authors proposed the transcriptional factor NFATc3 as the link between TRPC5 channels and MDR1 expression *(69)*. However we hypothesize that TRPC5 channels can also activate the PI3K/AKT pathway. In fact in another work it is shown the involvement of TRPC5 and TRPC6 channels in BDNF-induced AKT activation *(70)*, suggesting that, similar to Orai1 and Orai3 channels, TRPC5 and TRPC6 are the upstream activators of the PI3K/AKT pathway. We propose that TRPC5-mediated chemoresistance could be due to the activation of the PI3K/AKT pathway. In fact, a study shows that GSK3b inhibition increases translocation of NFATc3 to the nucleus *(71)*. Previously in this review we have shown that GSK3b inhibition was also associated with beta-catenin activation and subsequent c-myc expression. Because c-myc could be involved in MDR1 expression we propose that TRPC5 channels activate Akt. In his turn Akt inhibits GSK3b, resulting in both beta-catenin and NFATc3 translocation to the nucleus, MDR1 expression and subsequent chemoresistance. This hypothesis is consistent with the previous work in which authors propose NFATc3 as the link between TRPC5 channels and MDR1 expression *(69)*.

In another work similar mechanisms are described by authors *(72)*. Indeed, silencing TRPC5 with TRPC5 siRNA decreased MDR1 expression and increased chemosensitivity to 5-FU in HCT-8 and LoVo cells (colorectal cancer). Moreover silencing TRPC5 inhibited the beta-catenin pathway *(72)*. This result is consistent with our previous hypothesis that TRPC5 channels activate the beta-catenin pathway by inhibiting GSK3b.

These studies show that intracellular calcium concentration could play an important role in the development of chemoresistance.

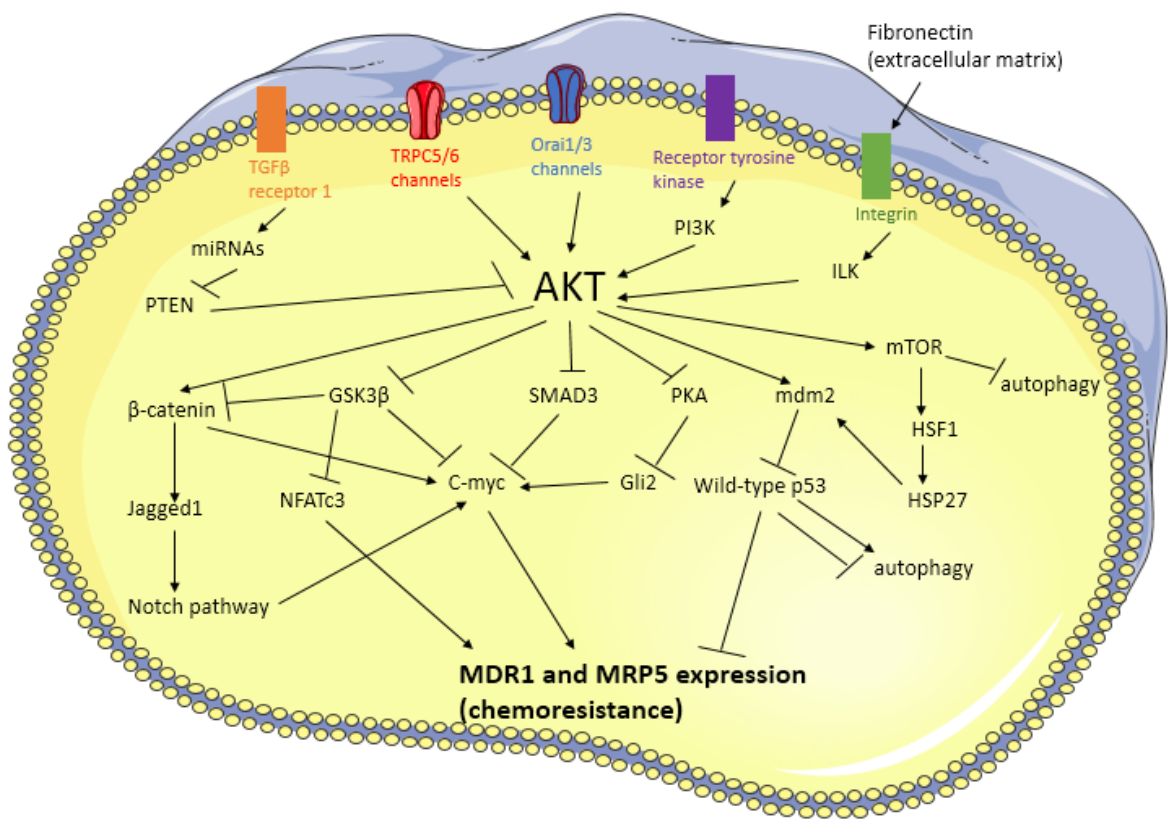

## Conclusion

We have describe the main pathways related to the Akt activation and involved in AKT-mediated chemoresistance in several cell lines. Furthermore we have shown that Akt could be activated by ion channels like Orai or TRPC channels. For this reason these channels could constitute an interesting therapeutic target in cancer to improve the effects of classical chemotherapeutic agents described in this review.

In this work we have described ABC transporters-dependent chemoresistance. However other mechanisms are responsible for chemoresistance like Bcl-xl with an anti-apoptotic effect for example *(20)*. In the same way, Akt appears to be the major regulator of p53 and c-myc but other mechanisms, not described in this review, can regulate p53 and c-myc. For example a recent study showed that p53 can be activated by the long non-coding RNA Meg3 *(61)*. This study also demonstrated that downregulation of Meg3 enhances cisplatin resistance in A549 cells through activation of the wnt/beta-catenin pathway, possibly via c-myc expression as described previously in this review. This result shows that long non-coding RNAs play important roles in chemoresistance and more generally in cancer progression. Another study showed that c-myc expression sensitizes medulloblastoma cells to radiotherapy and chemotherapy *(62)*. This result shows that c-myc can be associated with chemosensitivity to chemotherapy, suggesting that the role of c-myc in chemoresistance is very complex and may depend on cell lines. Finally, lithium chloride is used for bipolar disorder and it has been shown that lithium chloride inhibits GSK3b *(63)*. However as described previously in this review GSK3b could be involved in chemosensitivity to cisplatin in A549 cells by destabilizing oncoprotein c-myc and beta-catenin. For this reason we suggest that lithium chloride could increase chemoresistance to cisplatin in patients suffering from non small cell lung cancer and bipolar disorder. A previous study confirms this hypothesis because it has been shown that lithium chloride reduces the sensitivity of A549 cells to cisplatin and enhances nuclear translocation of beta-catenin *(20)*. This result shows the importance of the personalized medicine.

## References

1. Michael, R. Gold; Michael, P. Scheid; Lorna Santos; May Dang-Lawson; Richard A. Roth; Linda Matsuuchi; Vincent Duronio; Danielle L. Krebs. The B cell antigen receptor activates the Akt (protein kinase B)/Glycogen synthase kinase-3 signaling pathway via phosphatidylinositol 3-kinase. *J. Immunol.* **1999**, *163*, 1894-1905.
2. Ling-Zhi Liu, Xiang-Dong Zhou, Guisheng Qian, Xianglin Shi, Jing Fang and Bing-Hua Jiang. AKT1 Amplification Regulates Cisplatin Resistance in Human Lung Cancer Cells through the Mammalian Target of Rapamycin/p70S6K1 Pathway. *Cancer Res.* **2007**, 6325-32
3. Fraser M ; Bai T ; Tsang BK. Akt promotes cisplatin resistance in human ovarian cancer cells through inhibition of p53 phosphorylation and nuclear function. *Int J Cancer.* **2008**, 534-46
4. Abolfazl Avan ; Ravi Narayan ; Elisa Giovannetti ; Godefridus J Peters. Role of Akt signaling in resistance to DNA-targeted therapy. *World J Clin Oncol.* **2016**, 352-369
5. Zhang Y ; Bao C ; Mu Q ; Chen J ; Wang J ; Mi Y ; Sayari AJ ; Chen Y ; Guo M. Reversal of cisplatin resistance by inhibiting PI3K/Akt signal pathway in human lung cancer cells. *Neoplasma.* **2016**, 362-70
6. Xie C ; Pan Y ; Hao F ; Gao Y ; Liu Z ; Zhang X ; Xie L ; Jiang G ; Li Q ; Wang E. C-Myc participates in β-catenin-mediated drug resistance in A549/DDP lung adenocarcinoma cells. *APMIS.* **2014**, 1251-8

7. Yunching Chen ; Surendar Reddy Bathula ; Jun Li ; Leaf Huang. Multifunctional Nanoparticles Delivering Small Interfering RNA and Doxorubicin Overcome Drug Resistance in Cancer. *J Biol Chem.* **2010**, 22639-22650

8. Li A ; Song J ; Lai Q ; Liu B ; Wang H ; Xu Y ; Feng X ; Sun X ; Du Z. Hypermethylation of ATP-binding cassette B1 (ABCB1) multidrug resistance 1 (MDR1) is associated with cisplatin resistance in the A549 lung adenocarcinoma cell line. *Int J Exp Pathol.* **2016**

9. He Y ; Zhang J ; Zhang J ; Yuan Y. The role of c-myc in regulating mdr1 gene expression in tumor cell line KB. *Chin Med J (Engl).* **2000**, 848-51

10. Etienne Blanc ; David Goldschneider ; Eric Ferrandis ; Michel Barrois ; Gwenaëlle Le Roux ; Stéphane Leonce ; Sétha Douc-Rasy ; Jean Bénard ; Gilda Raguénez. MYCN Enhances P-gp/*MDR1* Gene Expression in the Human Metastatic Neuroblastoma IGR-N-91 Model. *Am J Pathol.* **2003**, 321-331

11. Gregory MA ; Qi Y ; Hann SR. Phosphorylation by glycogen synthase kinase-3 controls c-myc proteolysis and subnuclear localization. *J Biol Chem.* **2003**

12. Julienne R. Escamilla-Powers ; Rosalie C. Sears. A Conserved Pathway That Controls c-Myc Protein Stability through Opposing Phosphorylation Events Occurs in Yeast. *J Biol Chem.* **2007**, 5432-5442

13. Janardhan Sampath ; Daxi Sun ; Vincent J. Kidd ; Jose Grenet ; Amisha Gandhi ; Linda H. Shapiro ; Qingjian Wang ; Gerard P. Zambetti ; John D. Schuetz. Mutant p53 Cooperates with ETS and Selectively Up-regulates Human MDR1 Not MRP1. *J Biol Chem.* **2001**, 39359-39367

14. Robert A. Johnson ; Tan A. Ince ; Kathleen W. Scotto. Transcriptional Repression by p53 through Direct Binding to a Novel DNA Element. *J Biol Chem.* **2001**, 27716-27720

15. Masato Matsuoka ; Hideki Igisu ; Yasuo Morimoto. Phosphorylation of p53 protein in A549 human pulmonary epithelial cells exposed to asbestos fibers. *Environ Health Perspect.* **2003**, 509-512

16. Lindsey D. Mayo ; David B. Donner. A phosphatidylinositol 3-kinase/Akt pathway promotes translocation of Mdm2 from the cytoplasm to the nucleus. *Proc Natl Acad Sci USA.* **2001**, 11598-11603

17. Dexing Fang ; David Hawke ; Yanhua Zheng ; Yan Xia ; Jill Meisenhelder ; Heinz Nika ; Gordon B. Mills ; Ryuji Kobayashi ; Tony Hunter ; Zhimin Lu. Phosphorylation of β-Catenin by AKT Promotes β-Catenin Transcriptional Activity. *J Biol Chem.* **2007**, 11221-11229

18. Luo K ; Gu X ; Liu J ; Zeng G ; Peng L ; Huang H ; Jiang M ; Yang P ; Li M ; Yang Y ; Wang Y ; Peng Q ; Zhu L ; Zhang K. Inhibition of disheveled-2 resensitizes cisplatin-resistant lung cancer cells through down-regulating Wnt/β-catenin signaling. *Exp Cell Res.* **2016**, 105-13

19. Bush JA ; Li G. Cancer chemoresistance: the relationship between p53 and multidrug transporters. *Int J Cancer.* **2002**, 323-30

20. Zhang J ; Liu J ; Li H ; Wang J. β-Catenin signaling pathway regulates cisplatin resistance in lung adenocarcinoma cells by upregulating Bcl-xl. *Mol Med Rep.* **2016**, 2543-51

21. Fukumoto S ; Hsieh CM ; Maemura K ; Layne MD ; Yet SF ; Lee KH ; Matsui T ; Rosenzweig A ; Taylor WG ; Rubin JS ; Perrella MA ; Lee ME. Akt participation in the Wnt signaling pathway through Dishevelled. *J Biol Chem.* **2001**, 17479-83

22. Gao W ; Liu Y ; Qin R ; Liu D ; Feng Q.
Silence of fibronectin 1 increases cisplatin sensitivity of non-small cell lung cancer cell line. *Biochem Biophys Res Commun.* **2016**, 35-41

23. Rallis C ; Pinchin SM ; Ish-Horowicz D. Cell-autonomous integrin control of Wnt and Notch signalling during somitogenesis. *Development.* **2010**, 3591-601

24. Christina Cho ; Carol Horzempa ; David Jones ; Paula J. McKeown-Longo. The fibronectin III-1 domain activates a PI3-Kinase/Akt signaling pathway leading to αvβ5 integrin activation and TRAIL resistance in human lung cancer cells. *BMC Cancer.* **2016**

25. Delcommenne M ; Tan C ; Gray V ; Rue L ; Woodgett J ; Dedhar S. Phosphoinositide-3-OH kinase-dependent regulation of glycogen synthase kinase 3 and protein kinase B/AKT by the integrin-linked kinase. *Proc Natl Acad Sci USA.* **1998**, 11211-6

26. Estrach S ; Ambler CA ; Lo Celso C ; Hozumi K ; Watt FM. Jagged 1 is a beta-catenin target gene required for ectopic hair follicle formation in adult epidermis. *Development.* **2006**, 4427-38

27. Zhiwei Wang ; Yiwei Li ; Aamir Ahmad ; Asfar S Azmi ; Sanjeev Banerjee ; Dejuan Kong ; Fazlul H Sarkar. Targeting Notch signaling pathway to overcome drug-resistance for cancer therapy. *Biochim Biophys Acta.* **2010**, 258-267

28. Liu ZJ ; Shirakawa T ; Li Y ; Soma A ; Oka M ; Dotto GP ; Fairman RM ; Velazquez OC ; Herlyn M. Regulation of Notch1 and Dll4 by vascular endothelial growth factor in arterial endothelial cells: implications for modulating arteriogenesis and angiogenesis. *Mol Cell Biol.* **2003**, 14-25

29. Hannigan GE ; Leung-Hagesteijn C ; Fitz-Gibbon L ; Coppolino MG ; Radeva G ; Filmus J ; Bell JC ; Dedhar S. Regulation of cell adhesion and anchorage-dependent growth by a new beta 1-integrin-linked protein kinase. *Nature.* **1996**

30. Calzavara E ; Chiaramonte R ; Cesana D ; Basile A ; Sherbet GV ; Comi P. Reciprocal regulation of Notch and PI3K/Akt signalling in T-ALL cells in vitro. *J Cell Biochem.* **2008**, 1405-12

31. Giroux Leprieur E ; Vieira T ; Antoine M ; Rozensztajn N ; Rabbe N ; Ruppert AM ; Lavole A ; Cadranel J ; Wislez M. Sonic Hedgehog Pathway Activation Is Associated With Resistance to Platinum-Based Chemotherapy in Advanced Non-Small-Cell Lung Carcinoma. *Clin Lung Cancer.* **2016**, 301-8

32. Han L ; Pan Y ; Wang B. Small ubiquitin-like Modifier (SUMO) modification inhibits GLI2 protein transcriptional activity in vitro and in vivo. *J Biol Chem.* **2012**, 20483-9

33. Pawel Niewiadomski ; Jennifer H. Kong ; Robert Ahrends ; Yan Ma ; Eric W. Humke ; Sohini Khan ; Mary N. Teruel ; Bennett G. Novitch ; Rajat Rohatgi. Gli protein activity is controlled by multi-site phosphorylation in vertebrate Hedgehog signaling. *Cell Rep.* **2014**, 168-181

34. Pan Y ; Wang C ; Wang B. Phosphorylation of Gli2 by protein kinase A is required for Gli2 processing and degradation and the Sonic Hedgehog-regulated mouse development. *Dev Biol.* **2009**, 177-89

35. Kumar K ; Raza SS ; Knab LM ; Chow CR ; Kwok B ; Bentrem DJ ; Popovic R ; Ebine K ; Licht JD ; Munshi HG. GLI2-dependent c-MYC upregulation mediates resistance of pancreatic cancer cells to the BET bromodomain inhibitor JQ1. *Sci Rep.* **2015**

36. Wakefield LM ; Roberts AB. TGF-beta signaling: positive and negative effects on tumorigenesis. *Curr Opin Genet Dev.* **2002**, 22-9

37. Tang B ; Vu M ; Booker T ; Santner SJ ; Miller FR ; Anver MR ; Wakefield LM. TGF-beta switches from tumor suppressor to prometastatic factor in a model of breast cancer progression. *J Clin Invest.* **2003**, 1116-24

38. Remy I ; Montmarquette A ; Michnick SW. PKB/Akt modulates TGF-beta signalling through a direct interaction with Smad3. *Nat Cell Biol.* **2004**, 358-65

39. Wang X ; Chen X ; Meng Q ; Jing H ; Lu H ; Yang Y ; Cai L ; Zhao Y. MiR-181b regulates cisplatin chemosensitivity and metastasis by targeting TGFβR1/Smad signaling pathway in NSCLC. *Sci Rep.* **2015**

40. Kato M ; Putta S ; Wang M ; Yuan H ; Lanting L ; Nair I ; Gunn A ; Nakagawa Y ; Shimano H ; Todorov I ; Rossi JJ ; Natarajan R. TGF-beta activates Akt kinase through a microRNA-dependent amplifying circuit targeting PTEN. *Nat Cell Biol.* **2009**, 881-9

41. Yang L ; Wang YL ; Liu S ; Zhang PP ; Chen Z ; Liu M ; Tang H. miR-181b promotes cell proliferation and reduces apoptosis by repressing the expression of adenylyl cyclase 9 (AC9) in cervical cancer cells. *FEBS Lett.* **2014**, 124-30

42. Xia Y ; Gao Y. MicroRNA-181b promotes ovarian cancer cell growth and invasion by targeting LATS2. *Biochem Biophys Res Commun.* **2014**, 446-51

43. Sochor M ; Basova P ; Pesta M ; Dusilkova N ; Bartos J ; Burda P ; Pospisil V ; Stopka T. Oncogenic microRNAs: miR-155, miR-19a, miR-181b, and miR-24 enable monitoring of early breast cancer in serum. *BMC Cancer.* **2014**

44. Sun W ; Ma Y ; Chen P ; Wang D. MicroRNA-10a silencing reverses cisplatin resistance in the A549/cisplatin human lung cancer cell line via the transforming growth factor-β/Smad2/STAT3/STAT5 pathway. *Mol Med Rep.* **2015**, 3854-9

45. Lim SC ; Duong HQ ; Choi JE ; Parajuli KR ; Kang HS ; Han SI. Implication of PI3K-dependent HSP27 and p53 expression in mild heat shock-triggered switch of metabolic stress-induced necrosis to apoptosis in A549 cells. *Int J Oncol.* **2010**, 387-93

46. Xu Y ; Diao Y ; Qi S ; Pan X ; Wang Q ; Xin Y ; Cao X ; Ruan J ; Zhao Z ; Luo L ; Liu C ; Yin Z. Phosphorylated Hsp27 activates ATM-dependent p53 signaling and mediates the resistance of MCF-7 cells to doxorubicin-induced apoptosis. *Cell Signal.* **2013**, 1176-85

47. Lu H ; Sun C ; Zhou T ; Zhou B ; Guo E ; Shan W ; Xia M ; Li K ; Weng D ; Meng L ; Xu X ; Hu J ; Ma D ; Chen G. HSP27 Knockdown Increases Cytoplasmic p21 and Cisplatin Sensitivity in Ovarian Carcinoma Cells. *Oncol Res.* **2016**, 119-28

48. Lelj-Garolla B ; Kumano M ; Beraldi E ; Nappi L ; Rocchi P ; Ionescu DN ; Fazli L ; Zoubeidi A ; Gleave ME. Hsp27 Inhibition with OGX-427 Sensitizes Non-Small Cell Lung Cancer Cells to Erlotinib and Chemotherapy. *Mol Cancer Ther.* **2015**, 1107-16

49. Shiuh-Dih Chou ; Thomas Prince ; Jianlin Gong ; Stuart K. Calderwood. mTOR Is Essential for the Proteotoxic Stress Response, HSF1 Activation and Heat Shock Protein Synthesis. *PLoS One.* **2012**

50. Bhaskar PT ; Hay N. The two TORCs and Akt. *Dev Cell.* **2007**, 487-502

51. Bo Chen ; Yaoxi Tan ; Yan Liang ; Yan Li ; Lei Chen ; Shuangshuang Wu ; Wei Xu ; Yan Wang ; Weihong Zhao ; Jianging Wu. Per2 participates in AKT-mediated drug resistance in A549/DDP lung adenocarcinoma cells. *Oncol Lett.* **2017**, 423-428

52. Levine B ; Abrams J. p53: The Janus of autophagy? *Nat Cell Biol.* **2008**, 637-9

53. Wu T ; Wang MC ; Jing L ; Liu ZY ; Guo H ; Liu Y ; Bai YY ; Cheng YZ ; Nan KJ ; Liang X. Autophagy facilitates lung adenocarcinoma resistance to cisplatin treatment by activation of AMPK/mTOR signaling pathway. *Drug Des Devel Ther.* **2015**, 6421-31

54. Anne-Sophie Ay ; Nazim Benzerdjeb ; Henri Sevestre ; Ahmed Ahidouch ; Halima Ouadid-Ahidouch. Orai3 Constitutes a Native Store-Operated Calcium Entry That Regulates Non Small Cell Lung Adenocarcinoma Cell Proliferation. *PLoS One.* **2013**

55. Faouzi M ; Kischel P ; Hague F ; Ahidouch A ; Benzerdjeb N ; Sevestre H ; Penner R ; Ouadid-Ahidouch H. ORAI3 silencing alters cell proliferation and cell cycle progression via c-myc pathway in breast cancer cells. *Biochim Biophys Acta.* **2013**, 752-60

56. Wang M ; Liu ZM ; Li XC ; Yao YT ; Yin ZX. Activation of ERK1/2 and Akt is associated with cisplatin resistance in human lung cancer cells. *J Chemother.* **2013**, 162-9

57. Yagi K ; Furuhashi M ; Aoki H ; Goto D ; Kuwano H ; Sugamura K ; Miyazono K ; Kato M. c-myc is a downstream target of the Smad pathway. *J Biol Chem.* **2002**, 854-61

58. Zhao Q ; Gu X ; Zhang C ; Lu Q ; Chen H ; Xu L. Blocking M2 muscarinic receptor signaling inhibits tumor growth and reverses epithelial-mesenchymal transition (EMT) in non-small cell lung cancer (NSCLC). *Cancer Biol Ther.* **2015**, 634-43

59. Xu ZP ; Devillier P ; Xu GN ; Qi H ; Zhu L ; Zhou W ; Hou LN ; Tang YB ; Yang K ; Yu ZH ; Chen HZ ; Cui YY. TNF-α-induced CXCL8 production by A549 cells: Involvement of the non-neuronal cholinergic system. *Pharmacol Res.* **2013**, 16-23

60. Feng M ; Grice DM ; Faddy HM ; Nguyen N ; Leitch S ; Wang Y ; Muend S ; Kenny PA ; Sukumar S ; Roberts-Thomson SJ ; Monteith GR ; Rao R. Store-independent activation of Orai1 by SPCA2 in mammary tumors. *Cell.* **2010**, 84-98

61. Xia Y ; He Z ; Liu B ; Wang P ; Chen Y. Downregulation of Meg3 enhances cisplatin resistance of lung cancer cells through activation of the WNT/β-catenin signaling pathway. *Mol Med Rep.* **2015**, 4530-7

62. Von Bueren AO ; Oehler C ; Shalaby T ; von Hoff K ; Pruschy M ; Seifert B ; Gerber NU ; Warmuth-Metz M ; Stearns D ; Eberhart CG ; Kortmann RD ; Rutkowski S ; Grotzer MA. c-MYC expression sensitizes medulloblastoma cells to radio- and chemotherapy and has no impact on response in medulloblastoma patients. *BMC Cancer.* **2011**

63. Laure Freland ; Jean-Martin Beaulieu. Inhibition of GSK3 by lithium, from single molecules to signaling networks. *Front Mol Neurosci.* **2012**

64. Hu B ; Tack DC ; Liu T ; Wu Z ; Ullenbruch MR ; Phan SH. Role of Smad3 in the regulation of rat telomerase reverse transcriptase by TGFbeta. *Oncogene.* **2006**, 1030-41

65. Tang BD ; Xia X ; Lv XF ; Yu BX ; Yuan JN ; Mai XY ; Shang JY ; Zhou JG ; Liang SJ ; Pang RP. Inhibition of Orai1-mediated Ca2+ entry enhances chemosensitivity of HepG2 hepatocarcinoma cells to 5-fluorouracil. *J Cell Mol Med.* **2017**, 904-915

66. Selvaraj S ; Sun Y ; Sukumaran P ; Singh BB. Resveratrol activates autophagic cell death in prostate cancer cells via downregulation of STIM1 and the mTOR pathway. *Mol Carcinog.* **2016**, 818-31

67. Kim J ; Kundu M ; Viollet B ; Guan KL. AMPK and mTOR regulate autophagy through direct phosphorylation of Ulk1. *Nat Cell Biol.* **2011**, 132-41

68. Ma X ; Chen Z ; Hua D ; He D ; Wang L ; Zhang P ; Wang J ; Cai Y ; Gao C ; Zhang X ; Zhang F ; Wang T ; Hong T ; Jin L ; Qi X ; Chen S ; Gu X ; Yang D ; Pan Q ; Zhu Y ; Chen Y ; Chen D ; Jiang L ; Han X ; Zhang Y ; Jin J ; Yao X. Essential role for TrpC5-containing extracellular vesicles in breast cancer with chemotherapeutic resistance. *Proc Natl Acad Sci U S A.* **2014**, 6389-94

69. Ma X ; Cai Y ; He D ; Zou C ; Zhang P ; Lo CY ; Xu Z ; Chan FL ; Yu S ; Chen Y ; Zhu R ; Lei J ; Jin J ; Yao X. Transient receptor potential channel TRPC5 is essential for P-glycoprotein induction in drug-resistant cancer cells. *Proc Natl Acad Sci U S A.* **2012**, 16282-7

70. Fortin DA ; Srivastava T ; Dwarakanath D ; Pierre P ; Nygaard S ; Derkach VA ; Soderling TR. Brain-derived neurotrophic factor activation of CaM-kinase kinase via transient receptor potential canonical channels induces the translation and synaptic incorporation of GluA1-containing calcium-permeable AMPA receptors. *J Neurosci.* **2012**, 8127-37

71. Gómez-Sintes R ; Lucas JJ. NFAT/Fas signaling mediates the neuronal apoptosis and motor side effects of GSK-3 inhibition in a mouse model of lithium therapy. *J Clin Invest.* **2010**, 2432-45

72. Wang T ; Chen Z ; Zhu Y ; Pan Q ; Liu Y ; Qi X ; Jin L ; Jin J ; Ma X ; Hua D. Inhibition of transient receptor potential channel 5 reverses 5-Fluorouracil resistance in human colorectal cancer cells. *J Biol Chem.* **2015**, 448-56

73. Naruji K ; Arata N ; Tohru H ; Koji U ; Tadahiko E ; Tao-Sheng L ; Kimikazu H. The c-MYC-ABCB5 axis plays a pivotal role in 5-fluorouracil resistance in human colon cancer cells. *J Cell Mol Med.* **2015**, 1569-1581

74. Wakamatsu T ; Nakahashi Y ; Hachimine D ; Seki T ; Okazaki K. The combination of glycyrrhizin and lamivudine can reverse the cisplatin resistance in hepatocellular carcinoma cells through inhibition of multidrug resistance-associated proteins. *Int J Oncol.* **2007**, 1465-72

75. Wolfgang H ; Ralf F ; Martina S ; Matthias L ; Ralf J. Membrane Drug Transporters and Chemoresistance in Human Pancreatic Carcinoma. *Cancers (Basel).* **2011**, 106-125

76. Wang Q ; Beck WT. Transcriptional suppression of multidrug resistance-associated protein (MRP) gene expression by wild-type p53. *Cancer Res.* **1998**, 5762-9

77. Shanshan S ; Krista N.J ; Kimberly M.M ; Sekhar P.R ; Anne E.C ; Haiyang T ; Steven M.D ; Stephen M.B ; Joe G.N.G ; Ayako M ; Jason X.-J.Y. ATP promotes cell survival via regulation of cytosolic [Ca2+] and Bcl-2/Bax ratio in lung cancer cells. *Am J Physiol Cell Physiol.* **2016**, C99-C114